\documentclass[global,twocolumn,referee]{svjour}
\usepackage{graphicx}
\journalname{Appl. Phys. B}

\def\ignore#1{}
\def\fignore#1{}

\def\figbox#1{\Blue\fbox{#1} \Black}
\def\figbox#1{}

\def\be{\begin{equation}}
\def\ee{\end{equation}}
\def\bea{\begin{eqnarray}}
\def\eea{\end{eqnarray}}

\def\Blue{}

\def\Black{}

\def\ignore#1{}

\graphicspath{{figures\ }}
\begin{document}
\title{Ranging with frequency-shifted feedback lasers:
from  $\mu$m-range accuracy to MHz-range measurement rate
}
\author{J. I. Kim \inst{1,5} \and V. V. Ogurtsov \inst{1,4} \and G. Bonnet \inst{2} \and L. P. Yatsenko \inst{3} \and K. Bergmann \inst{1,6}
}                     
\institute{ Department of Physics, Technical University Kaiserslautern, 67653 Kaiserslautern,
Germany
\and
SPHERON-VR AG, Hauptstrasse 186, 67714 Waldfischbach-Burgalben, Germany
\and
Institute of Physics, National Academy of Sciences of Ukraine, prospect Nauki 46, Kiev-39,
03650, Ukraine, E-mail: leonid.yatsenko@gmail.com
 \and
present address: Zollamstr. 45, 67663 Kaiserslautern, Germany
\and
present address: Defense R\&D Center, Hanwha Corporation, 99 Oesam-ro, Yuseong-gu,
Daejeon 305-156, Korea
\and
OPTIMAS research center, Technical University Kaiserslautern, 67653 Kaiserslautern,
Germany
 }

\date{Received: date / Revised version: date}
%
\maketitle
\begin{abstract}
We report results on ranging based on frequency shifted feedback (FSF) lasers with two different implementations:   (1) An Ytterbium-fiber system for measurements  in an  industrial environment with accuracy of the order of 1  $\mu$m, achievable over a distance of the order of meters with potential to reach an accuracy of better than 100 nm; (2) A semiconductor laser system for a high rate of measurements  with  an accuracy of 2 mm @ 1 MHz or 75  $\mu$m @ 1 kHz and a limit of the accuracy of    $\geq $ 10  $\mu$m. In both implementations, the distances information is derived from a frequency measurement. The method is therefore insensitive to detrimental influence of ambient light. For the Ytterbium-fiber system a key feature is the injection of a   single frequency laser, phase modulated at variable frequency  $\Omega$, into the FSF-laser cavity. The frequency  $\Omega_{max}$ at which the detector signal is maximal yields the distance. The semiconductor FSF laser system operates without external injection seeding. In this case the key feature is frequency counting  that allows convenient choice  of either   accuracy or speed of measurements  simply by changing the duration of the interval during which the frequency is measured by counting.
\end{abstract}
\section{Introduction}
\label{intro}
Following up on earlier work in theory \cite{YAT04b,YAT09b} and experiment \cite{OGU06a,OGU08} we report new results using the concept of ranging based on a frequency-shifted feedback laser (FSF-laser). Central to these concepts is the use of a frequency comb.  Therefore our  work appears fitting for this volume honoring T. H\"{a}nsch at the occasion of his 75$^{\rm th}$ birthday. However, the frequency comb used in our work is created by a cavity-internal acousto-optical frequency shifter (AOFS) and thus differs from the type of optical comb for which T. H\"{a}nsch received the Nobel Prize \cite{HAE06}. Our work is directed towards the development of systems which are suitable for industrial applications, i.e. which are accurate, robust and economic.

 In Sect.~\ref{sec:2} we present work based on a fiber-laser system aiming at high accuracy of the measurement over distances of the order of meter. An accuracy  $\delta z  \approx 1$  $\mu$m is documented ($\delta z/z \approx 10^{-6}$), with an expected limit of the achievable accuracy $<$ 100 nm ($\delta z/z < 10^{-7}$).

 In Sect.~\ref{sec:3} we present work based on a semiconductor laser aiming at a high rate of measurements. We document the potential to achieve a measurement rate up to 1 MHz with mm accuracy.  At a lower   measurement rate, the accuracy of measurements with that system approaches 10  $\mu$m.

\subsection{The FSF laser concept}
\label{introA}

The history of work on lasers with cavity-internal frequency shift goes back many decades, at least to the early 1970, when a Stanford group \cite{STR70} later including also T. H\"{a}nsch \cite{TAY71} showed that insertion of an AOFS into the laser cavity allowed  fast wavelength tuning. That work led   to the development of a new type of modelocking: Kowalski et. al. \cite{KOW87} showed that robust modelocking occurs when the AOFS frequency shift equals the free spectral range of the cavity. This was demonstrated first in a passive  cavity \cite{KOW87} and later with an active cavity \cite{KOW88b,BON96}. The group at Golden \cite{KOW88a,HAL90} was also the first to start characterizing the optical output spectrum of a FSF laser.

The latter work triggered similar experimental and theoretical efforts in Kaiserslautern \cite{LIT91a,LIT92,BAL93,KOW94} and later in Sendai \cite{NAK97,NAK98}. In parallel, the group of T. H\"{a}nsch continued to further develop the scheme as a means for tuning laser wavelength \cite{RIC91a} or bandwidth control \cite{RIC91b}. Aspects of the nonlinear dynamics of a FSF laser were analyzed via simulation studies in \cite{SAB94,STE03,MAR06} or in experiments \cite{MAR94}. More recent work looking at the properties of FSF lasers can be found e.g. for fiber lasers doped with Erbium \cite{VAZ13,VAZ14} or Thulium \cite{CHE16}, or for semiconductor lasers in \cite{NDI08,SEL15}.

A comprehensive general theory of the FSF-laser operation was first presented in \cite{YAT04a} followed by work with special emphasis on the coherence properties of the optical output \cite{YAT09a}. Further detailed analysis of FSF-laser properties and some new applications was presented by the Grenoble group \cite{CHA09,CHA10,CHA11,CHA13b,PIQ13,CHA14,CHA16}.

\subsection{Other applications of FSF lasers}
\label{introB}
Although our present work treats \emph{ranging} based on FSF lasers, we briefly mention a few other applications of such lasers. Early work using specific properties of the FSF laser concerned the mechanical action of light on atoms, such as  slowing and cooling of atoms in a beam \cite{LIT91b} following up on a proposal by \cite{HOF88}  or light-induced drift separation of Rb isotopes \cite{MUG93}. Further applications include  accurate frequency-interval measurements \cite{BAR92};  efficient optical pumping of atoms \cite{LIM98,CAS00}; efficient excitation of atoms in the higher atmosphere for preparation of a guide star \cite{PIQ03,MAR07,PIQ11};   observation of dark resonances in optical excitation of atoms \cite{ROM10};   realization of a FSF laser  with high \cite{BAL95} or ultrahigh and variable \cite{CHA13a} pulse rate;   control of pulse duration \cite{NIK09} or new schemes for frequency stabilization \cite{NIK10}; rapid tuning of frequency \cite{LYA16}; and   generation of a frequency comb for specific tasks in underwater sensing \cite{ZHA16}.

\subsection{	Laser-based distance measurements}
\label{introC}

Numerous  methods for laser-based distance measurement are known. Several  of them are implemented in commercial devices.  Here we only mention a few reviews that cover many of the well known techniques at various levels of detail \cite{AMA01,GRO01,GEN11,BER12a,DEL12}. An interesting line of more recent development is based on H\"{a}nsch-type frequency combs, often with sub-mm accuracy, some of them developed for applications over very  long distances in space (e.g. \cite{COD09,LEE10,BER12b,BAU14,LIA14,JAN15,BER15,JAN16}). One common feature  of most of the techniques discussed in those reviews is the need to  accurately measure, in one   way  or another, intensities returned from the object. Ambient light reaching the detector may therefore be detrimental to the achievement of high sensitivity or accuracy. To conclude this brief overview of past work, we mention early work  \cite{WEB88} suggesting   ranging based on interferometry with phase modulated light. Because this approach
relies on quantitative measurement of light intensity-modulation amplitude it has not found widespread application for ranging.

\subsection{		Ranging based on FSF lasers}
\label{introD}

An early application of a FSF laser for distance measurement was reported by the Sendai group \cite{NAK00} using a Nd-YAG laser followed by later work \cite{ITO08,NDI09} including the application to the monitoring of bridge movements over 100~m range distances \cite{UME10}. A powerful modification of the FSF-ranging scheme was introduced by the Kaiserslautern group through theoretical work \cite{YAT04b} followed by experimental demonstrations using lasers with an Ytterbium-doped fiber \cite{OGU06a,OGU06b} or
Erbium-doped fiber \cite{OGU08} as gain medium.
 The theoretical frame of that work is discussed in some detail in Sect. \ref{sec:2A}. Further contributions to the scheme came from \cite{SHO06} and in particular from the Grenoble group \cite{PIQ13}.  Applications using a variety of gain media, e.g. Titanium-Sapphire \cite{BRA10} and distributed feedback systems \cite{HON07a,HON07b} have also been reported.

A significant feature that sets FSF laser-based ranging technique apart from more traditional methods is the fact that the distance of an object is deduced from a frequency measurement.

\begin{figure*}
\resizebox{0.9\textwidth}{!}{
   \includegraphics[angle=90]{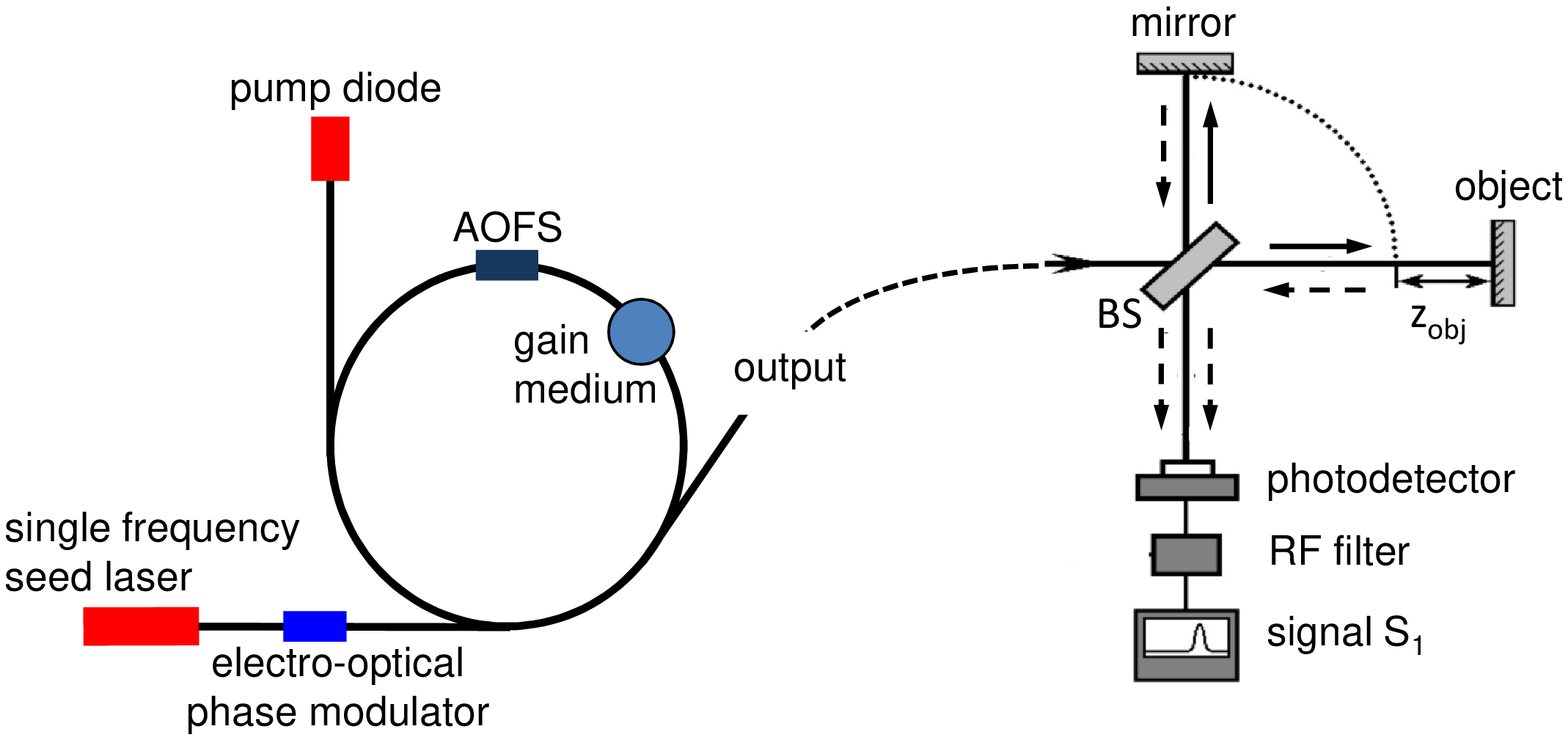}
  }
\caption{On the left side, the schematics
 of the FSF ring-laser is shown, with a Ytterbium-doped fiber as gain medium, an acousto-optical frequency-shifter AOFS, a pump diode and a single frequency ($\approx 1$ MHz bandwidth) seed laser  injected into the fiber ring after passing an electro-optical phase modulator. The laser output is launched into the ranging setup, shown at the right side, with reference mirror, object, beam splitter (BS), photodetector, RF-filter and evaluation electronics.  The central frequency of the RF-filter coincides with the variable  frequency $\Omega(t)$ of phase modulation.}
\label{fig1}       
\end{figure*}
%

\section{The fiber laser  system: overview}
\label{sec:2}

\subsection{The Ytterbium  system}
\label{sec:2.1}

The characteristic elements of the FSF laser used for this experiment are shown in Fig.\ref{fig1}. The essential components are a broadband gain medium (here a Yb$^{3+}$-doped optical fiber), an intracavity acousto-optical frequency shifting element (AOFS), a pump laser (diode laser) and an output port. The AOFS is inserted into the ring via fiber pigtails adjusted such that only the frequency-shifted first diffraction order of the AOFS is allowed to continue circulating in the ring cavity. A specific and crucial feature of this setup is a  single-frequency CW laser, the radiation of which is injected into the ring cavity after passing through an electro-optical phase modulator \cite{BER03}.

The radiation reflected from the reference surface and the time-delayed comb scattered  from the object are superimposed on the detector leading to a multi-component signal, see Eqs. (\ref{eq9a}) and (\ref{eq9b})  below. The follow-up electronics looks at the amplitude of the ac-component of the signal at frequency $\Omega$. The modulation frequency  $\Omega_0$ at which this signal is maximal, yields -- according to Eq. (\ref{eq13}) below -- directly the distance to the object. Thus, a measurement proceeds by tuning the modulation frequency $\Omega$  over a suitable frequency range. During the scan the frequency $\Omega_0$ is determined.

Ranging with an accuracy of the order of 1  $\mu$m requires a broad optical spectrum which can be realized by e.g. a fiber laser system as used in this experiment.  Yb$^{3+}$-doped optical fibers  have a large gain bandwidth due to strongly broadened laser transitions in glasses. Ytterbium ions Yb$^{3+}$ have a    simple electronic level structure, with only one excited state manifold ($^2$F$_{5/2}$) which can be exited   from the ground-state manifold ($^2$F$_{7/2}$) with near-infrared radiation. The level scheme has   a quasi-three-level character. Compared to the semiconductor gain medium (see Sect. \ref{sec:3}) Yb$^{3+}$-doped fiber gain has a much larger saturation intensity  and longer upper-state lifetime (typically of the order of 1--2 ms).

All components are spliced together leaving no adjustable elements.  Figure~\ref{fig1} shows also the schematics of the Michelson arrangement implemented for detecting the ranging signal. An extra mirror for the reference path may not be needed if the reflection from the end facet of the fiber, which delivers the radiation to the object, serves as reference beam.

\subsection{Comparison with FMCW-Ladar}
\label{sec:2.2}
At first glance the FSF-laser ranging concept appears similar to the frequency-modulated scheme for laser detection and ranging (FMCW-Ladar, see e.g. \cite{BER12a,DIE94,MAT15}), a technique which allows measuring a beat frequency   resulting from   single-frequency   laser radiation,  the frequency of which is linearly changed in time with a chirp rate  $\gamma_c$ periodically repeated in a saw-tooth pattern. That radiation is sent via a beam splitter to a reference surface and the object before being recombined at the detector. Any difference $ z_{obj}$  in path length to the reference mirror and the object results in a beat frequency $\Omega_{beat}$  from which $
 z_{obj}$ is deduced according to
\be
                                                   z_{obj}= \frac{c}{2} \frac{\Omega_{beat}}{\gamma_c} .
                                                  \label{eq1}
\ee
Obviously for high accuracy a large  $\Omega_{beat}$  and thus a large $\gamma_c$   (a steep chirp) is wanted. Furthermore, high accuracy requires a strictly linear variation of frequency with time.

However, the physics underlying  the generation of a ranging signal by a FSF laser  differs from the physics of FMCW-Ladar, see Sect. \ref{sec:2A}.

\section{Theoretical background }
\label{sec:2A}

\subsection{The FSF-laser field }

\label{sec:2A1}

Following \cite{YAT04a} the output field ${E} (t)$ of the FSF laser, seeded by laser radiation of amplitude $\mathcal{E}_0$, frequency  $\omega_s$ and phase $\varphi_s(t)$, is a sum of many discrete components
\be
{E} (t)=\sum_{n=0}^{\infty} \mathcal{E}_n \exp[-i(\omega_s+ n\Delta)t -i\Phi_n-i\varphi_s(t -n\tau )],
\label{eq3}
\ee
because the seed laser is strong enough to dominate over any seeding by spontaneous emission within the gain medium.  The frequency of component $n$ is shifted from $\omega_s$ by $n\Delta$ where $\Delta$ is the AOFS-induced frequency shift per round trip. The phase of this component is given by the phase $\varphi_s$ of the seed laser taken at the time $t_0= t-n\tau$ (which is the time when the radiation relevant for this component entered the cavity), augmented by the additional phase shift  $\Phi_n$ due to the propagation within the cavity,
\be
  \Phi_n = -n \tau [\omega_s + (n+1) \Delta_{} / 2 ] ,
  \label{eq4}
  \ee
where  $\tau$ is  the cavity round-trip  of the FSF laser. Saturated gain from the amplifying medium and loss from any spectral filter that may act in the cavity alter the original seed  amplitude  $\mathcal{E}_0$ to the values  $\mathcal{E}_n$. The distribution of amplitudes $\mathcal{E}_n$ of the discrete spectral components is well described \cite{YAT04a} by a Gaussian

 \be
  \mathcal{E}_n = \mathcal{E}_0 \exp \left[ - \frac {(n -
n_{max})^2}{n_W^2} \right].
\label{eq5}
  \ee
  centered at the component $n_{max}$ where the intensity reaches its maximum. The parameter $\Delta  n_{max}$ is the shift of the maximum of the optical spectrum of the FSF laser   from the frequency of the seed laser. The parameter $n_{W}$ is the width of the Gaussian distribution of amplitudes $\mathcal{E}_n$. Accordingly, the width of the optical  spectrum is determined by the distribution of intensities $\propto |\mathcal{E}_n|^2$ and is equal to $\Delta n_{W}/\sqrt{2}$.   Typically  we have $n_{max} \sim n_{W}$ and $10^3 <n_{W}<10^4$.

  The phase of the seed laser is periodically changed by an electro-optical modulator (EOM) according to
\be
 \varphi_s(t) = \beta \sin (\Omega t + \vartheta),
 \label{eq6}
 \ee
where  $\Omega$ is the modulation frequency,  $\beta$   the modulation index and  $\vartheta$   an (irrelevant) initial phase. The frequency $\Omega$  of this phase modulation is a crucial parameter in the experiment.

\subsection{The structure of the ranging signal}
\label{sec:2A2}

The fields returned from the reference surface and   from   the object are combined at the detector. The signal delivered by the detector depends on $t$, $\Omega$  and $T$, where $T$ is the delay of the time of arrival of radiation coming from the reference surface and  from  the object. The signal is evaluated from
\be
S(t;\Omega  , T) = \left\langle\left|{E}(t;\Omega ) + E(t-T;\Omega )\right|^2\right\rangle,
\label{eq7}
\ee
where  $\left\langle\,\, \right\rangle$ means averaging over a period long compared to the optical cycle $\omega_s^{-1}$ but short compared with the modulation cycle $ \Omega^{-1}$.

The signal $S(t; \Omega , T)$ is a superposition of many harmonics of the modulation frequency \cite{YAT04b}
\be
S(t; \Omega , T) =\sum_{l} S_l(\Omega , T) \exp(il\Omega t) .
\label{eq8}
\ee
We use an electronic   filter to restrict detection of the signal at the modulation frequency $\Omega$ yielding  $S_1(\Omega , T)$. Using Eqs. (\ref{eq3}), (\ref{eq4}), (\ref{eq6}), and (\ref{eq8}) we obtain
\be
S_1(\Omega , T) =   J_1(X) \sum_n | \mathcal{E}_n|^2\left[P_n^{-} -
 P_n^{+}\right]
\label{eq9a}
\ee
with
\be
P_n^{(\pm)} =\exp\{-i[n (\Omega\tau\pm T\Delta) -\psi ]\}.
\label{eq9b}
\ee
Here $J_1(X)$ is a Bessel function and  $X= 2\beta\sin(\Omega T/2)$. The phase
   \be
   \psi=\omega_s T - (\Omega T + \pi )/2 + \theta
   \label{eq10}
   \ee
does not depend on $n$. The term $n\Omega\tau$  originates from the propagation  within the FSF cavity and $nT\Delta$ results from   the accumulated AOFS-induced shift of the optical frequency of the $n$-th component of the field, Eq. (\ref{eq3}).

In the experiment the signal $|S_1(\Omega , T)|$ is measured as a function of $\Omega$. For each $n$ we have two contributions to $S_1(\Omega , T)$ with amplitude according to Eq. (\ref{eq5}). The phase factors $P_n^{(\pm)}$, Eq. (\ref{eq9b}), depend on $n$, the system parameters  $\tau$ and $\Delta$ and the variables $\Omega$  and $T$. As $n$ increases from $n = 1$ to $n > n_{max} $ ($ n_{max} \approx  10^4$)  the phases $n (\Omega\tau\pm   T\Delta)-\psi$    of $P_n^{(\pm)}$   will go, for given $\Omega$  and $T$, through very many cycles of $2\pi$  and thus their contribution to $|S_1( \Omega, T)|$ will be small unless we  have either $\Omega\tau-   T\Delta= 2\pi m$  or  $\Omega\tau+   T\Delta= 2\pi k$ with integer $m$ and $k$. In that case the phases of either the first term, associated with $P_n^{-}$, or the second term, associated with $P_n^{+}$, will be equal to $2\pi mn -\psi$ or $2\pi kn -\psi$, respectively, meaning that $P_n^{-}$ or $P_n^{+}$  is in effect independent of $n$.

Assuming $T>0$ (as in our experiment) we realize that for the modulation frequency  $\Omega=\Omega_q^{+}$ given by
\be
\Omega_q^{+}= \gamma_c T  + q\Delta_{FSR};\quad q = -M_0, -M_0+1,\ldots
\label{eq11}
\ee
the terms of Eq. (\ref{eq9a}) associated with $P_n^{+}$ tend to sum up to a very small value, if not zero. However, the coefficients $P_n^{-}$ are independent of $n$ and the sum reaches a possibly large absolute value $S_{max}$. In Eq. (\ref{eq11})  $\gamma_c= \Delta/\tau $ is the chirp rate in the "moving comb" model of FSF laser \cite{YAT04a,KAS98}, and $M_0$ is the  integer part of $T\Delta /2\pi$. Similarly, for    the modulation frequency  $\Omega=\Omega_p^{-}$  given by
 \be
\Omega_p^{-}= -\gamma_c T  + p\Delta_{FSR};\quad p = M_0+1, M_0+2,\ldots
\label{eq12}
\ee
the terms of Eq. (\ref{eq9a}) associated with $P_n^{-}$ tend to sum up to a very small value. In this case, the coefficients $P_n^{+}$  are independent of $n$ and the absolute value of the sum reaches the same large value $S_{max}$.

The electronics for the data analysis tracks the detector signal $|S_1(\Omega , T)|$, for a given $T = 2  z_{obj}/c$. Here $ z_{obj} = 0$ marks the location of the end of the fiber,  where the reference laser beam and the beam sent to the object are separated. In the present experiment this is the end facet of the optical fiber which launches the laser beam towards the object.  The modulation frequency  $\Omega_{max}$ at which $|S_1( \Omega , T)|$ reaches its maximum is recorded.  For a short distance $ T <2\pi /\Delta$ (meaning  $M_0=0$), as is typical for our experiment,  the first two frequency  $\Omega_{max} $ are  $\Omega_0^{+} = \gamma_c T $    and $\Omega_1^{-}= -\gamma_c T+\Delta_{FSR}$. We can use both of these frequencies to find the distance $  z_{obj}$. In the present experiment we use the resonance at $\Omega_0^{+} = \gamma_c T $ and find the distance $ z_{obj}$ from
\be
z_{obj}=  \frac{c}{2}  \frac{\Omega_0^{+} }{\gamma_c}.                          \label{eq13}
\ee

When the measurement aims at an accuracy of 1 $\mu$m the order $q$ or $p$  will usually be known. When this is not the case the ambiguity can be resolved experimentally, see Sect. \ref{sec:3}.

In case of a smooth dependence of the amplitudes $\mathcal{E}_ n$ on $n $ the function $S_1(\Omega , T)$, Eq. (\ref{eq9a}), represents the Fourier transform of the optical spectrum of the FSF laser. For a Gaussian optical spectrum, Eq. (\ref{eq5}), the variation of the $|S_1(\Omega , T)|$ with  $\delta\Omega=\Omega-\Omega_{max}$ is also Gaussian:
\be
|S_1(\Omega , T)| = S_{max}\exp\left\{-\left[\frac{\delta\Omega^2}{\delta\Omega_0^2}\right]\right\},    \label{eq15}
\ee
where
\be
 \delta\Omega_0 = \frac{2\sqrt{2}}{n_W\tau} .
 \label{eq16}
 \ee
The method is not sensitive to detrimental influence of ambient light or to a variation of the intensity of the light returned from the object, as long as $|S_1(\Omega , T)|$ is large enough to reliably determine  $\Omega_{max}$.

Furthermore the accuracy  $\delta z $ of a measurement is determined by the width  $\delta\Omega_0$ of     $|S_1(\Omega , T)|$. The accuracy  $\delta \Omega_{max}$ of a measurement of $\Omega_{max}$ is  $\delta \Omega_{max}=\epsilon \delta \Omega_0$ where $\epsilon$  is determined by the electronic recording system and laser noise. In our experiment we typically have $10^{-3} < \epsilon  < 10^{-2}$. According to Eqs. (\ref{eq5}), (\ref{eq13}) and (\ref{eq16}), the uncertainty of $\Omega_{max}$ leads to uncertainty  $\delta z$ of distance
\be
 \delta z = \epsilon \frac{2c}{\Gamma_{opt}},                                     \label{eq17}
 \ee
where $\Gamma_{ opt} =  n_w\Delta/\sqrt{2}$  is the width of the optical spectrum of the FSF laser. Thus the attainable accuracy of the distance measurements is given exclusively by the width of the optical spectrum and does \emph{not} depend on distance. Furthermore, because the signal $S_1(\Omega , T)$ is caused by beating the radiation which propagates along the reference path ($I_{ref}$) and the path to the object ($I_{obj}$), its amplitude is $ \propto \sqrt{I_{ref}I_{obj}}$ and thus varies with $ z_{obj}$ as $1/  z_{obj}$ rather than as $1/  z_{obj}^2$.

\section{Experiment}
\label{sec:2B}

\subsection{The fiber FSF laser}
\label{sec:2B1}

The all-fiber Yb$^{3+}$-doped FSF ring laser (schematically shown in Fig. \ref{fig1}) with a free-spectral range $\Delta_{FSR}=113,6$ MHz ($\tau=8.8$~ns)  is a compact transportable system similar to the one used in refs. \cite{OGU06a}, \cite{OGU08}, and \cite{OGU06b}. The gain medium of the laser is a 40 cm long Yb$^{3+}$-doped active fiber pumped by a diode laser  ($\lambda  =  976$~nm, $P = 300$~mW) which is coupled into the ring via  a wavelength division multiplexer (WDM). The intra-cavity fiber-pigtail acousto-optic frequency shifter  increases the frequency of the radiation circulating in the cavity by 200~MHz per round trip, leading to a "chirp rate" of   $\gamma_c = \Delta/\tau =2.3\cdot 10^{16}$ Hz/s. The single frequency seed laser is a diode laser in Littmann-Metcalf  configuration \cite{LIT78,LIU81}. The voltage driving the EOM phase modulator is generated by a digital synthesizer.

 \begin{figure}
 \centerline{\includegraphics[width=9cm,angle=0]{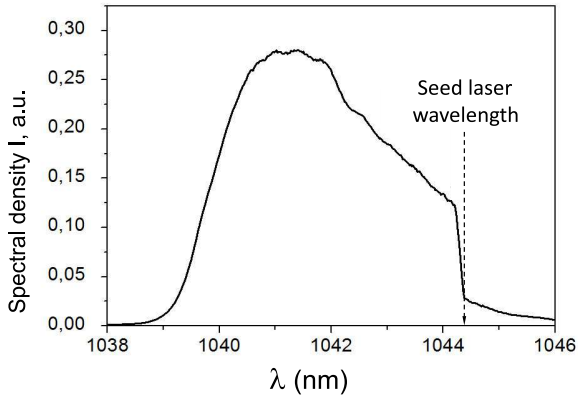}}%
 \caption{  The spectum of the FSF-laser radiation recorded with low resolution. The comb structure is not resolved. The seed-laser frequency is marked. The AOFS shift is to higher frequencies.
}
 \label{fig2}       
 \end{figure}

Figure \ref{fig2} shows a typical spectrum of the FSF-laser radiation with its central wavelength near   $\lambda_c = 1041$~nm and width   of 5~nm. Experiments are typically done with an output power of $P = 10$~mW. Fiber couplers are used to inject the seed  laser into the cavity and to couple   radiation out of the cavity. All fiber components are connected by fusion splices.

\subsection{The detection system}
\label{sec:2B2}

\begin{figure}[h]
\centering{\includegraphics[width=80mm]{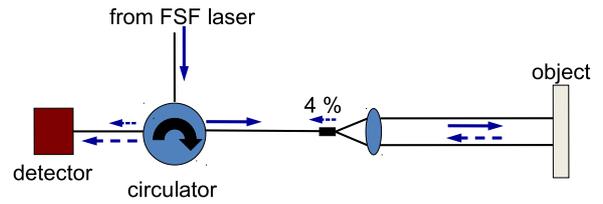}}%
 \caption{  Details of the Michelson interferometer setup. The circulator directs the radiation from the FSF laser  via a fiber to the object. The light reflected  from the end facet of the fiber serves as the reference beam. The reflected beams are directed by the circulator to the detector.}
 \label{fig3}
 \end{figure}

The modulation frequency $\Omega (t)$ is linearly scanned across an interval $\Delta \Omega$  which includes the resonance frequency according to Eq. (\ref{eq11}) or Eq. (\ref{eq12}). Typically $\Delta \Omega$  is set to cover a range of  changes of $  z_{obj}$  less than    2~mm. The FSF-laser radiation is sent via fibers through a circulator  before being launched towards the object. The reference beam is either generated by the reflection off the end facet of the fiber, see Fig. \ref{fig3}, or in a Michelson interferometer setup, see Fig. \ref{fig1}.

 The returning  radiation is coupled into the same fiber, reaches the circulator again and is directed by it to the fast InGaAs photodetector of area 0.2~mm$^2$  with fixed gain, a bandwidth of 150 MHz and a sensitivity range of 800--1700~nm.  The detector signal passes an electronically controlled  bandpass filter with a high transmission in the vicinity of the modulation frequency $\Omega$   before being sent to the evaluation electronics which records  the modulation frequency  $\Omega_{max}$ at which the signal $|S_1(\Omega , T)|$ reaches its maximum.  An automatic level-control assures that the evaluation electronics processes a signal of about the same magnitude, within limits independent of the intensity of the light returned from the object.

\subsection{Results}
\label{sec:2C}

\subsubsection{Accuracy}
\label{sec:2C1}
\begin{figure*}
 \centerline{\includegraphics[width=10cm,angle=90]{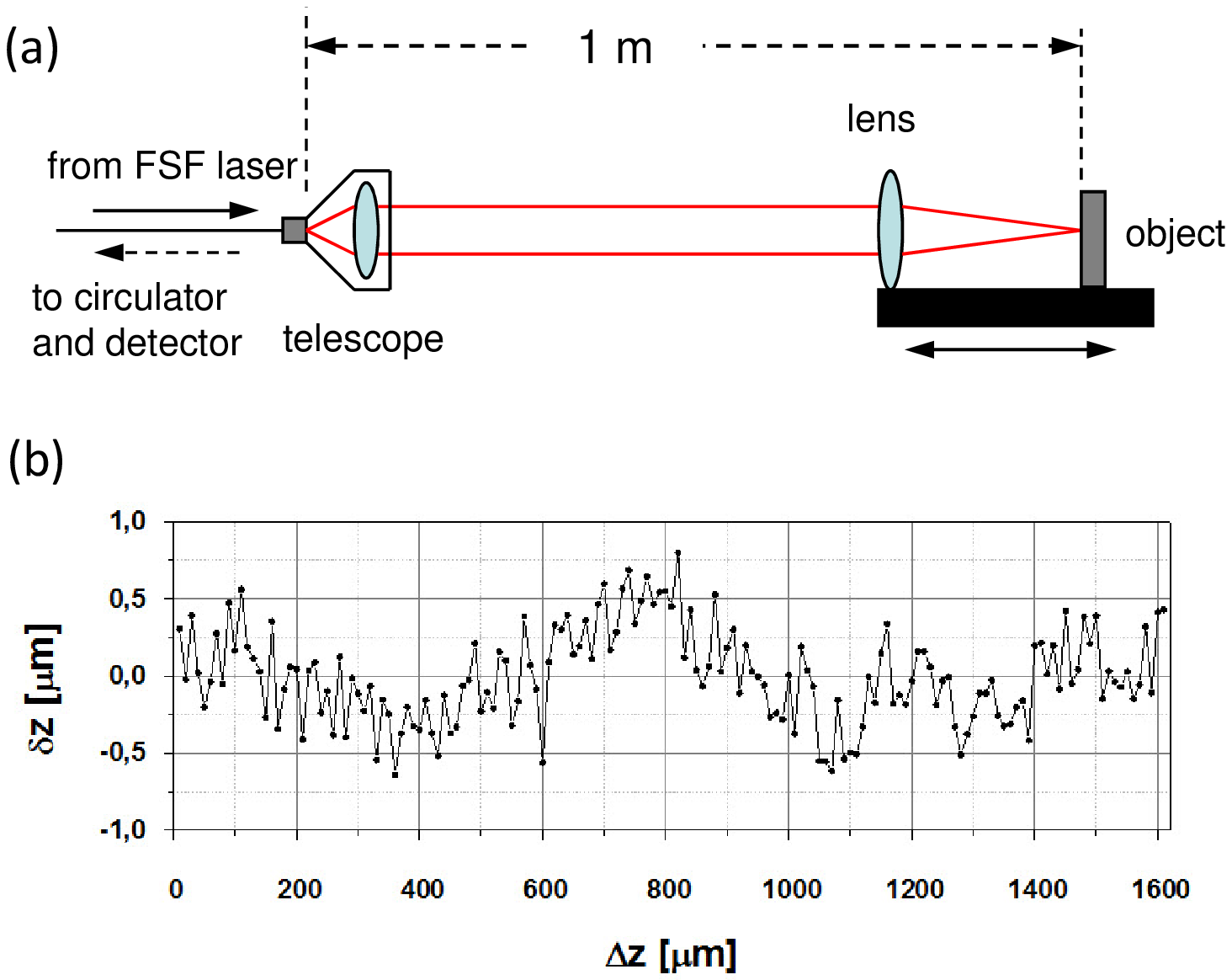}}%
 \caption{ Accuracy of the FSF laser based measurement. (a) Optical setup. The object and the focusing lens are jointly mounted on a platform. The position of the platform is varied under the control of a linear encoder with better than 100~nm accuracy. (b) Deviation $\delta z $ of the result provided by the linear encoder and by the FSF laser based measurement while the position of the platform is changed over a range of $\Delta z_{max}= 1.6$~mm. The position $\Delta z = 0 $ corresponds to $  z_{obj} = 1002,139$~mm, known from the frequency measurement to an accuracy of 1~$\mu$m.}
 \label{fig4}
 \end{figure*}

The data shown in Fig. \ref{fig4} are taken with the FSF-laser radiation launched towards the object via a telescope (Fig. \ref{fig4}a)
which expands the beam diameter to 8~mm before it is focused to a 15~$\mu$m spot onto the surface. For testing the accuracy of the system, the object and the focusing lens are both mounted on a platform which is moved to change the distance $  z_{obj}$.  Thus, the size of the laser spot on the object does not change when $  z_{obj}$ is varied. The variation of the position $  z_{obj}$ of the platform, and thus the position of the object, is independently monitored with a Heidenhain linear encoder   with an accuracy of better than 100~nm. An Abbe error may arise  from an angular deviation $\theta$ of the direction along which the platform motion is measured  from the direction of propagation and the   laser beam. Therefore the angle $\theta$ was kept below 0.1 degrees. In this case  the Abbe error does not  exceed 10 nm.

The accuracy of the laser-ranging result is documented in Fig. \ref{fig4}b with the object  placed at $  z_{obj}\approx   1$~m (the exact reading from the frequency measurement for $\Delta z=0$ is $  z_{obj} = 1002,129$~mm accurate to 1~$\mu$m) and moved in steps of 10~$\mu$m over the distance $\Delta z = 1.6$~mm while the variation of the position is measured through both the FSF-laser system and the linear decoder. Figure \ref{fig4}b shows the deviation  $\delta z$ of these two readings. Each point is an average over one hundred measurements, each   of 2~ms duration.   The deviation  $\delta z$ of the results of the laser measurement from the result of the linear encoder is  $\delta z \leq   0.8 $~$\mu$m  ($\delta z/z< 10^{-6}$). The apparently periodic deviation  from  $\delta z= 0$ by no more than 0.6~$\mu$m is caused by diffraction on the focusing lens which can   shift   $\Omega_{max}$ from the value given by Eq. (\ref{eq11}). An analysis [Yatsenko and Bergmann, to be published] shows that under uncontrolled conditions (nonideal focusing, titled object, non-Gaussian FSF laser spectrum) this deviation may reach the order of $\mu$m.

\begin{figure}
 \resizebox{0.5\textwidth}{!}{%
 \includegraphics[angle=90]{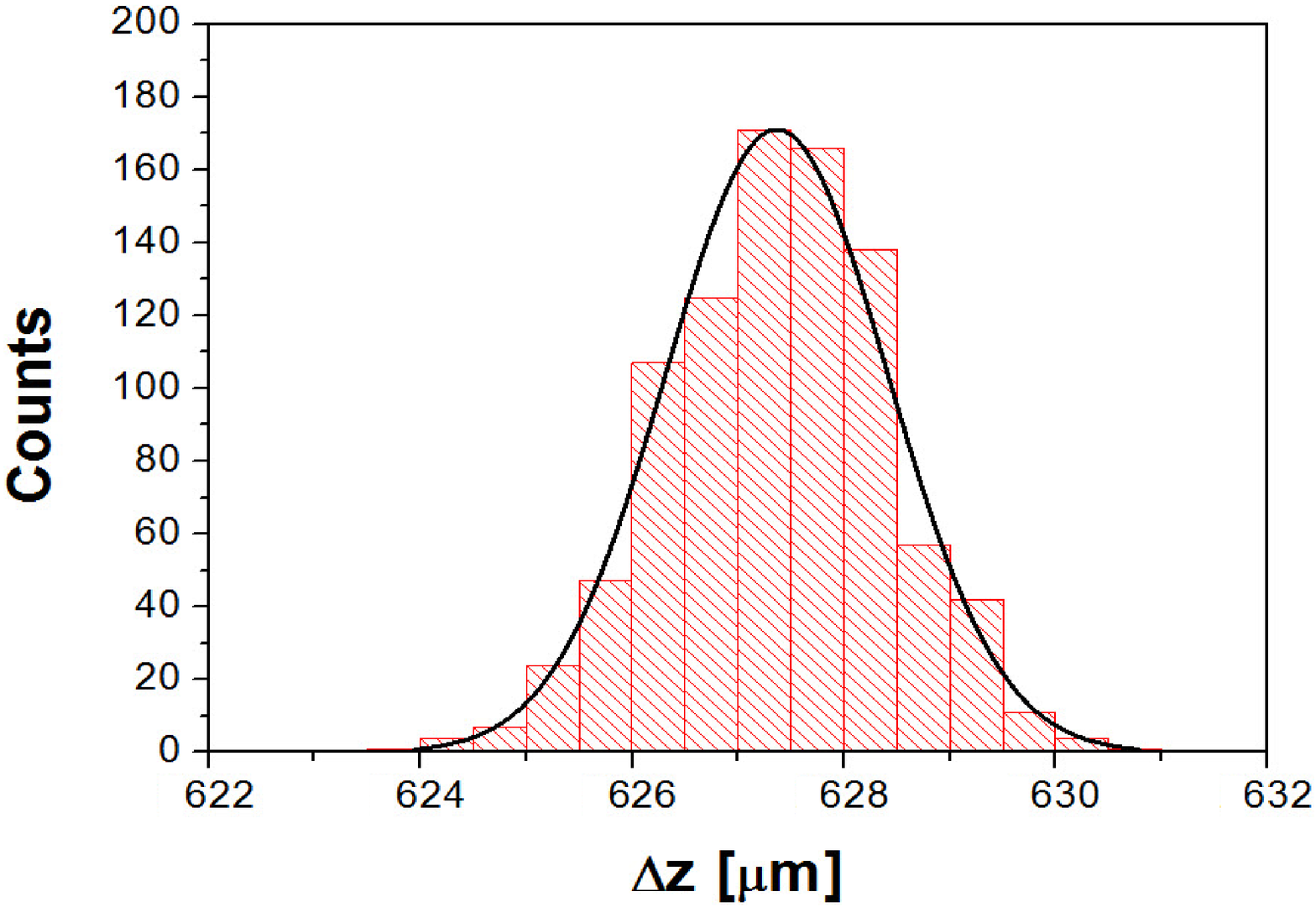}%
}
 \caption{ Distribution of the result of 10$^3$ FSF laser based measurements taken at a fixed position near $\Delta z = 627$~$\mu$m, see Fig. \ref{fig4}. A Gaussian fit leads for this distribution to the standard deviation of  $\sigma = 1.1$~$\mu$m.}
 \label{fig5}
 \end{figure}

Further characterization of the system is documented in Fig. \ref{fig5}. Here the distribution of results from 10$^3$ sweeps of $\Omega$  across the bandwidth    $\Delta \Omega$ are shown for a fixed distance $  z_{obj} = 1002,627$~mm. The distribution is characterized  by a standard deviation of $\sigma= 1.1$~$\mu$m. We emphasize that the width of the distribution is independent on the distance $  z_{obj}$ to the object. The accuracy of a single measurement, taken during a 2 ms sweep of  $\Omega$, is  $3\sigma \approx 3$~$\mu$m. If time is available for e.g. a thousand measurements (in a total of a few seconds) the maximum of the distribution, like the one shown in Fig. \ref{fig5},  can easily be determined to less than 10\% of its width corresponding to an uncertainty of 100 nm or less.
 The accuracy of the measurements   is determined by the width of the resonance  (see Eq. (\ref{eq17})).  Like most   other  absolute ranging methods the FSF laser ranging technique determines the optical path length between the laser and the object. Therefore, for larger distances, the variation of the refractive index along the path during the measurement could be important. Due to air turbulence, variation of the temperature profile etc. the air refractive index will deviate from the value $(n_{air}-1)10^{8}=27403.6$  \cite{EDL66}. According to  \cite{JAN16} the  relative uncertainty of the distance can be estimated to be about $\delta z_{obj}/z_{obj}=2\cdot 10^{-8}$. For our experimental conditions the influence of the uncertainty of $n_{air}$ needs to be considered only for $z_{obj}>5$~m.

\subsubsection{Ambiguity}
\label{sec:2C2new}

Like in any modulation-based ranging system,   FSF-ranging exhibits ambiguity regarding the results. A measured distance is known only modulo the addition of a multiple of the unambiguity range, which -- for our Ytterbium system -- is of the order of 1~m. The latter range is determined exclusively by the size of the AOFS frequency shift. However, being interested in micron-range accuracy   the prior uncertainty is usually  much less than the ambiguity range.  Should that not be the case, the procedure discussed in Sec. \ref{sec:3B} is applicable.

\subsubsection{Scanning object surfaces}
\label{sec:2C2}

A series of experiments  was conducted to demonstrate the high quality of surface topography recording. In this case, the focusing lens, shown in Fig. \ref{fig4}a, is not attached to the table. The scanning of the object in the $x-y$-plane is done with stepper motors in increments of 10~$\mu$m. The variation of the position is monitored by linear encoders of the type mentioned above.  The coordinate $z$ is measured through the FSF laser ranging method. The surface topography is reconstructed from the measured coordinates ($x,y,z$).

\begin{figure}
 \resizebox{0.40\textwidth}{!}{%
 \includegraphics[angle=90]{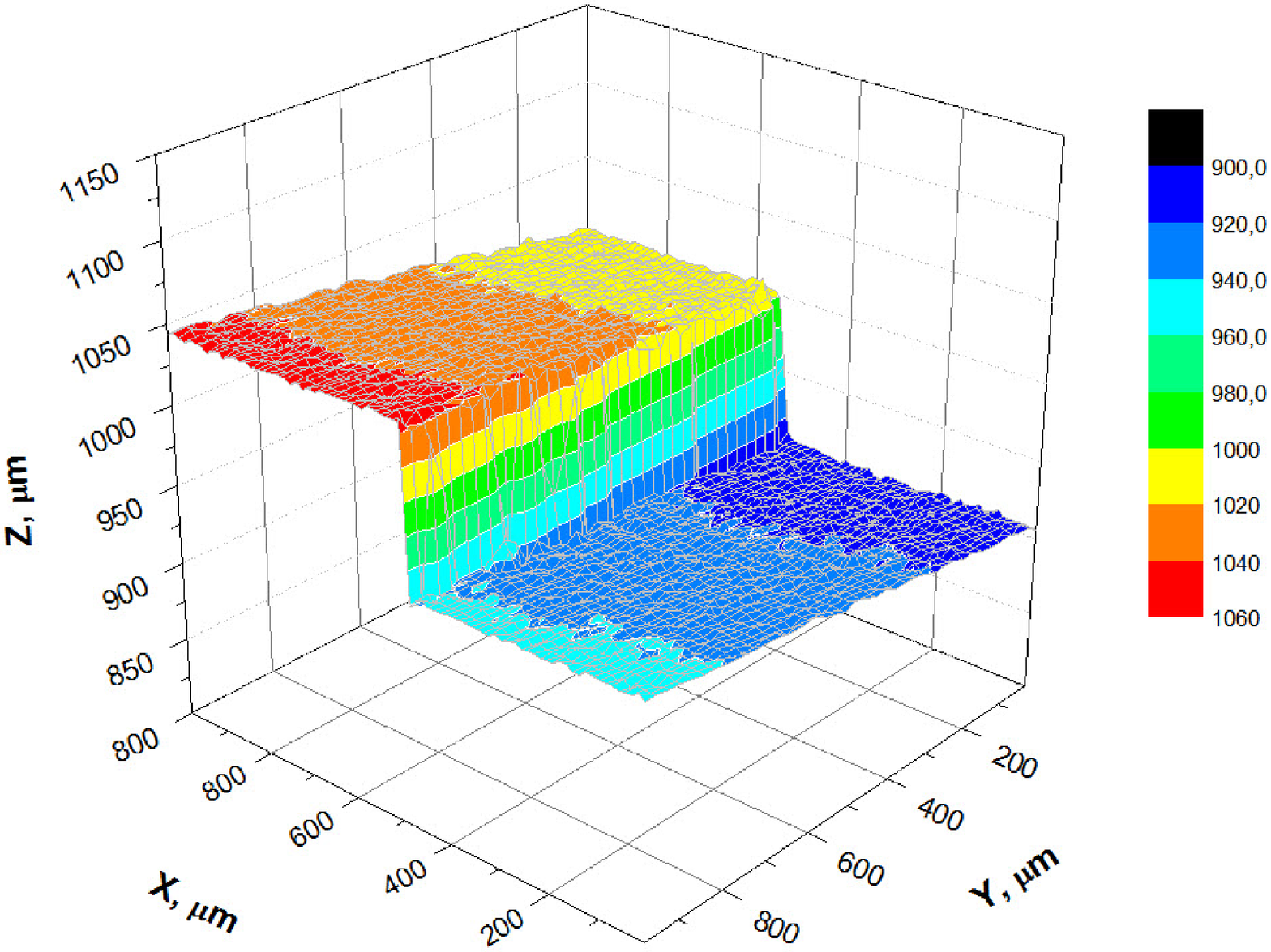}%
}
 \caption{ False colour representation of the topography of a flat black unpolished metal surface with a 110~$\mu$m step, taken from a distance of 1 m.  The surface normal was not exactly parallel to the axis of the laser beam. The step size in the $x$- and $y$-direction was 10~$\mu$m. The diameter of the laser spot on the surface was 15~$\mu$m. }
 \label{fig6}
 \end{figure}

Figure \ref{fig6} shows a step of 110~$\mu$m in an otherwise flat  blackened-metal surface positioned at a distance of $z_{obj} \simeq 1$~m. The surface normal is not oriented exactly parallel to the direction of the incident laser beam axis.  The average amplitude of the wiggles marking the boundary between colors is consistent with the accuracy of the order of 1~$\mu$m, as documented in Figs. \ref{fig4} and \ref{fig5}.

\begin{figure}
 \resizebox{0.37\textwidth}{!}{%
 \includegraphics[angle=90]{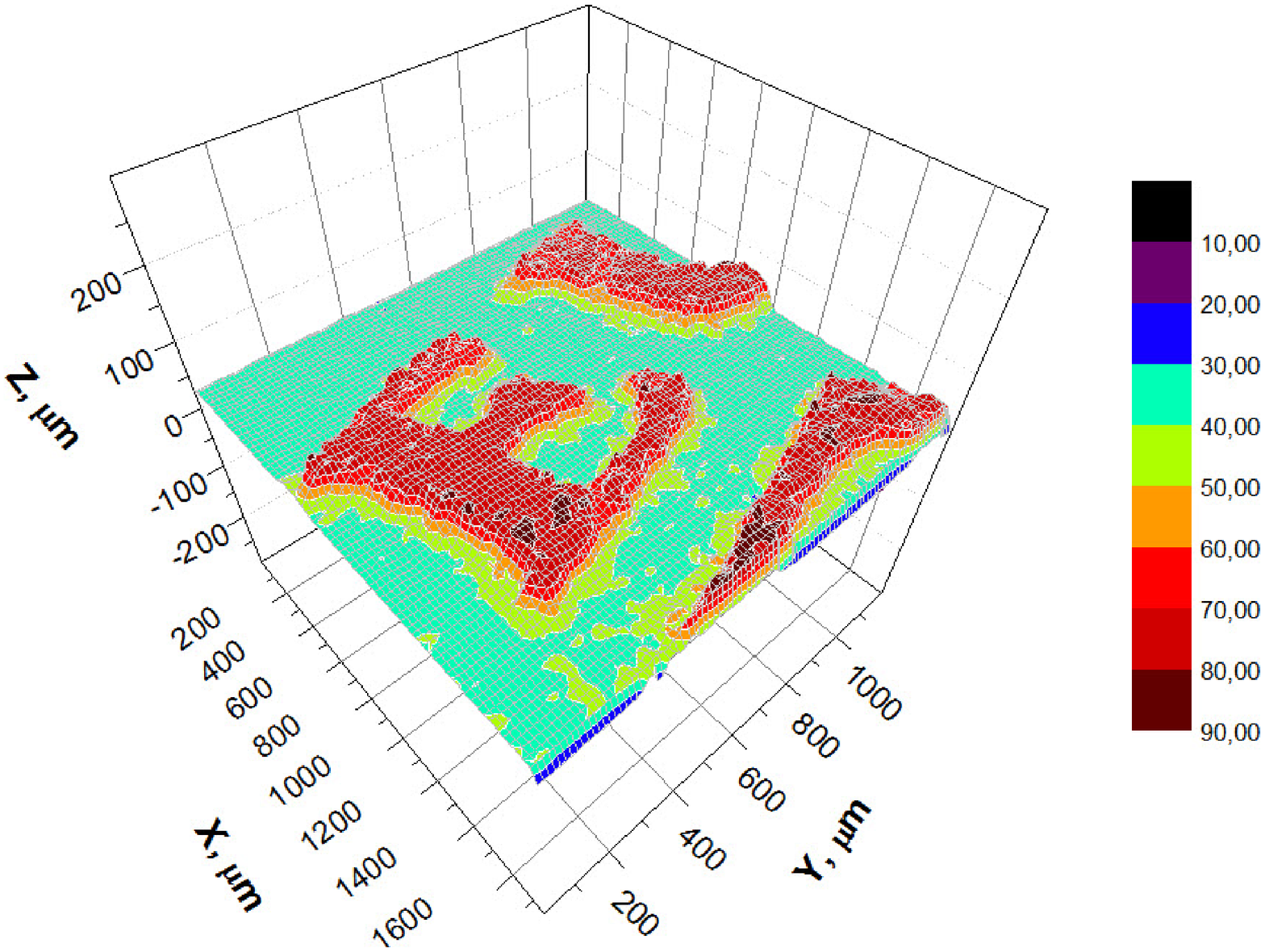}%
}
 \caption{ False colour representation of the surface topography of a segment from a Euro 2-cent coin taken from a distance of 1~m. The experimental conditions are the same as for Fig.~\ref{fig6}.
}
 \label{fig7}
 \end{figure}

\begin{figure}
 \resizebox{0.475\textwidth}{!}{%
 \includegraphics[angle=90]{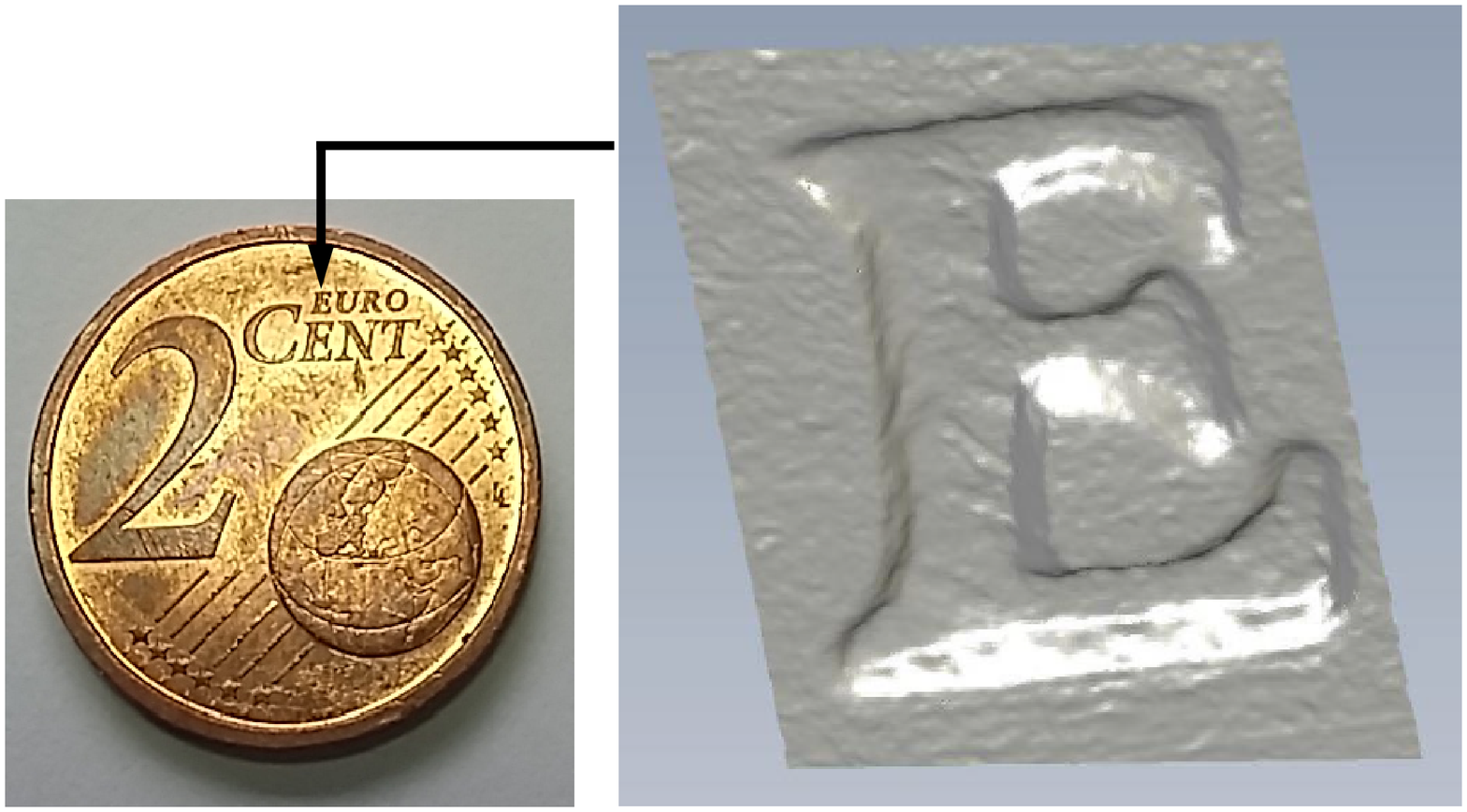}%
}
 \caption{ The right  frame shows the region of a Euro 2-cent coin (see left frame) near the letter E as recorded by FSF laser ranging at a distance of 1 m after rendering, based on the same set of data used for Fig.~\ref{fig7}.}
 \label{fig8}
 \end{figure}

Figure \ref{fig7}  shows the surface topography of the letter "E" of a Euro 2-cent coin (see Fig. \ref{fig8}) also taken from a distance of $ z_{obj} \simeq 1$~m. Figure \ref{fig8} shows the same letter "E", now using the full information content of the data set (see Fig. \ref{fig7}), after rendering, together with a photo of the coin. The high potential of our approach may be best appreciated by comparison with \cite{AFL15}, which shows the topography of an US one-cent coin recorded with a $4 \times 4$ array of detectors, using a time-domain frequency modulated continuous wave ranging scheme.

\section{The semiconductor-laser system}
\label{sec:3}

Many applications are anticipated (e.g. documentation of objects of meter-size or larger) where speed of measurement, rather than the highest possible accuracy is important. The minimum duration  $\Delta t_{min}$ of a measurement of the kind described in the previous section for a fiber-laser system is set by the width of the RF resonance. Recalling (see Eq. (\ref{eq5})) that $n_W =\sqrt{2} \Gamma_{ opt}/\Delta$, we find from Eq. (\ref{eq16})  with  $\Delta t_{min} = 1/\delta\Omega_0$
\be
 \Delta t_{min} = \frac{\Gamma_{opt}}{2\Delta}\tau.
 \label{tmeas}
 \ee	
Thus, for high speed measurements it is preferable to work with a smaller optical bandwidth, a large  $\Delta$  and in particular a short cavity. The lower limit for the length of the cavity is given by the size of the AOFS and the dimensions of the gain medium. When a semiconductor gain medium is used the size of the cavity can be much smaller compared to a system based on a doped fiber. In fact we have  $\tau= 0.8$~ns  ($\Delta_{FSR} = 1.3$~GHz) for the system discussed below.

However, some properties of a semiconductor gain medium differ distinctly from those of doped fibers. For instance, the upper state lifetime in a semiconductor system is orders of magnitudes shorter than in a fiber system and the rate of spontaneous emission is accordingly larger. It is therefore difficult, if not impossible, to control the emission spectrum of the FSF laser towards yielding a single frequency comb by injecting a narrow band laser as successfully done for the fiber system. As a consequence, we do not expect for a semiconductor gain medium the optical spectrum of the FSF laser to consist of a well-defined frequency comb with isolated frequency components separated by $\Delta$. Rather, each photon spontaneously emitted into a cavity mode will be amplified in successive round trips and create a comb with frequency components separated by $\Delta$. Because the spontaneous emission processes occur throughout the spectral profile the superposition of all the frequency combs started by individual emission processes will lead to a continuum spectrum with no immediately obvious comb structure.

\subsection{Theoretical background}
\label{sec:3A}

Without external seed laser, the output of the FSF laser is dominated by spontaneous emission processes within the gain medium. These processes are uncorrelated and intrinsically incoherent. However, the fourth-order cor\-re\-la\-tion-function analysis \cite{YAT09a}, which allows the evaluation of the RF spectrum  of the output intensity of a Michelson interferometer, shows that this spectrum comprises a set of doublets, whose frequencies are given by Eqs. (\ref{eq11}) and (\ref{eq12}). This structure in the RF spectrum of a Michelson-interferometer output results from the correlation of interference terms of individual components of a cyclostationary stochastic process \cite{YAT09a}. The spectrum carries information about the path length difference of the interferometer arms (the distance to be measured) and was used in early work for ranging \cite{NAK00}.   Of interest is the frequency for which the spectral density of the fluctuation reaches a maximum. This approach is similar to the one described for the fiber system in Sect.~\ref{sec:2}  but is less accurate because the origin of the resonance is noise \cite{YAT04a}.

Another option, used here, for accessing the information about distance is the direct frequency counting of the quasi-monochromatic output of the  RF filter. This approach requires the isolation of only a single resonance   in the RF spectrum of the interferometer output (see. Sect. \ref{sec:3B}).
This approach allows a  convenient choice of either  speed or  precision of the measurement by simply choosing an appropriate duration of the gate during which the frequency is determined by counting oscillation periods.

The interferometer output intensity filtered near one of the frequencies given by Eq. (\ref{eq11}) or Eq. (\ref{eq12}) is the quasi-monochromatic signal
\be
I(t)=\delta I(t)\cos [\Omega_{max} t+\phi(t)].
\label{eq18}
\ee
Here $\delta I(t)$ and $\phi(t)$ are slow random function of time with autocorrelation time $1/ \delta \Omega_0$ where   $ \delta \Omega_0$ is the RF resonance width given by Eq. (\ref{eq16}).  Theoretical analysis  [Yatsenko and Bergmann, to be published] shows that the mean value $\delta I_0=\sqrt{\langle \delta I^2(t)\rangle} $  of the amplitude of the oscillating part of the interferometer output depends on the width $\Gamma_{opt}$ of the optical spectrum of the FSF laser and is given by
\be
 \delta I_0 =2\sqrt{(1-R)R} |r|\frac{(2\pi)^{1/4}}{\sqrt{\Gamma_{opt}\tau}}  I_0 .
\label{eq19}
\ee
Here $R$ is the ratio of the power sent to the reference mirror and the object surface, the coefficient $r$ describes the efficiency of the collection system efficiency,  $I_0$ is the mean output intensity  of the FSF laser. The amplitude  $\delta I_0$ can be quite large especially for the semiconductor gain medium used for our experiment in combination with a short cavity length. For example, for $R=1/2$, $r=1$ (which is the case for a highly reflecting surface), $\Gamma_{opt}=200$~GHz,  $\tau=1$~ns we have $\delta I_0 \simeq 0.045 I_0$ which can be conveniently recorded and evaluated.

\begin{figure}
 \resizebox{0.5\textwidth}{!}{%
  \includegraphics[angle=0]{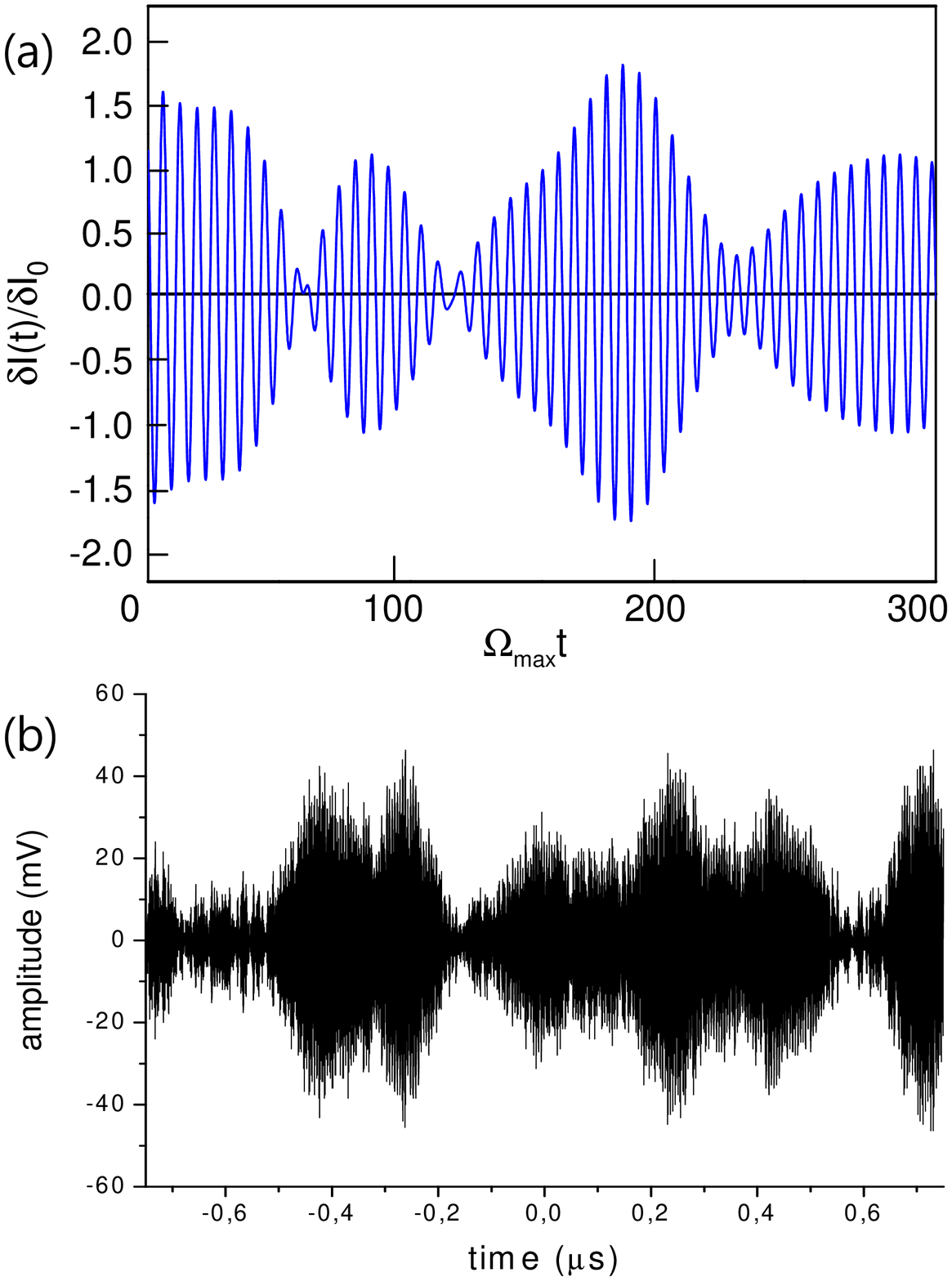} %
}
 \caption{(a) A simulated  realization of the noise process, Eq. (\ref{eq18}),  for a Gaussian spectrum of  width $\Gamma= 0.1 \Omega_{max}$.   (b)  Experimental trace of the oscillating component of  the    linear-cavity FSF-laser output (see Sect. \ref{sec:3c}).   }
 \label{fig9}
 \end{figure}

However, the noise origin of the oscillatory signal  leads to a challenge in the practical implementation of the scheme. Figure \ref{fig9} shows a simulated  realization of the process, Eq. (\ref{eq18}),  for a Gaussian spectrum   with    width  $\Gamma=0.1 \Omega_{max}$   and  an experimental trace of the oscillating component of  the    linear-cavity FSF-laser output described in Sect. \ref{sec:3c}.   The fluctuation of the amplitude does not prevent the frequency counting except when the signal amplitude is close to zero, as occurs at random times. The short period, during which the amplitude falls below a threshold needed for the counting must be excluded by electronic means to obtain the highest possible accuracy.

\subsection{ Elimination of ambiguity}

\label{sec:3B}

 Interferometric measurements determine the distance only to within an integer multiple of an unambiguity distance $z_{ua}$ \cite{OGU08}. Procedures which allow resolving the ambiguity of the detector reading are well known. The measures required in our case needs a discussion of the RF spectrum of the detector signal.

\begin{figure}[h]
\centerline{
 \includegraphics[width=8 cm,angle=0]{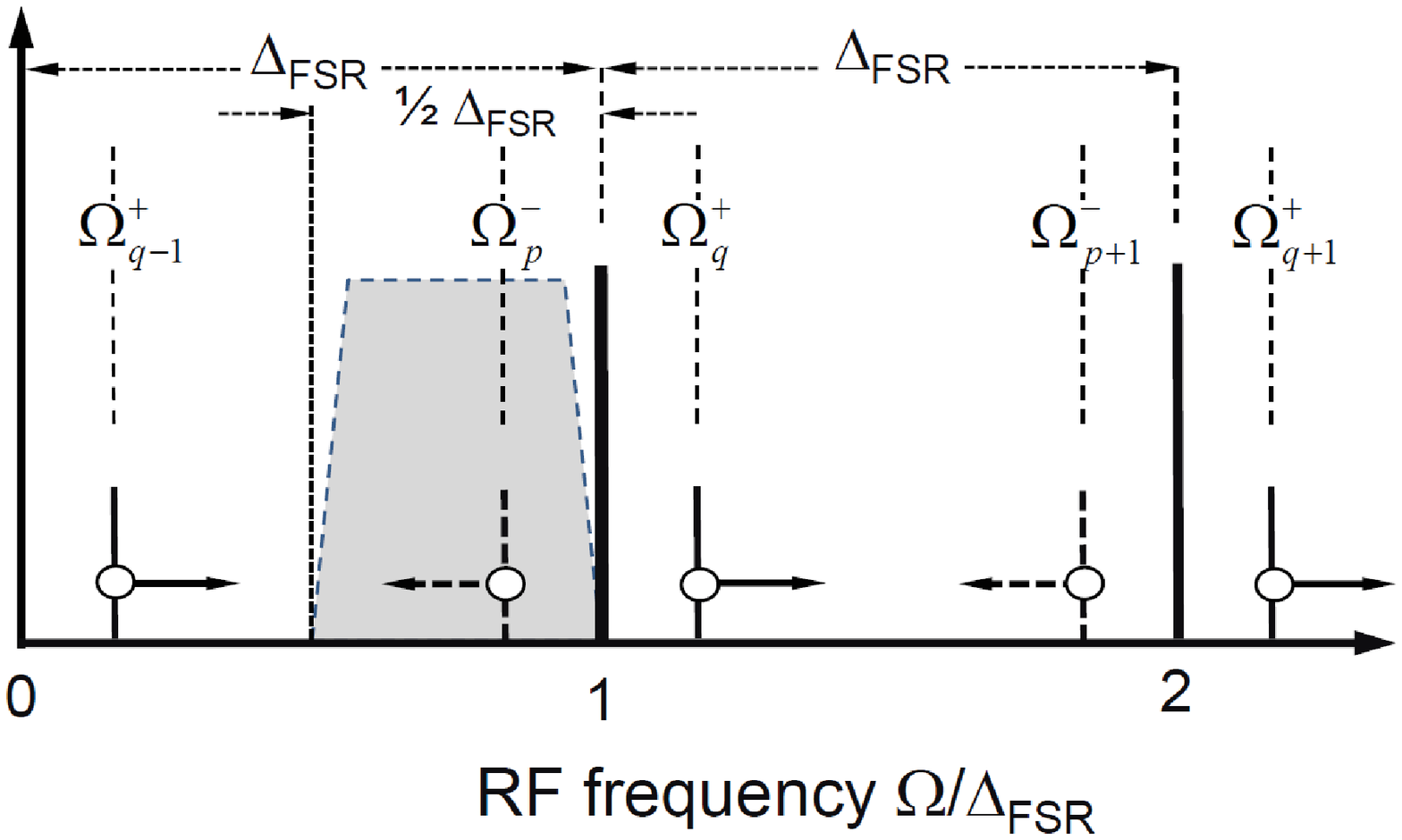}%
}
 \caption{  The  RF
spectrum recorded by the detector.
The frequency range to be selected for
detection and frequency counting is
marked by the trapezoidal shaded area. The arrows attached to the short dashed or solid lines
indicate the direction of change of their frequency
when $z_{obj}$ and thus $\gamma_c T$ increase. }
 \label{fig10}
 \end{figure}

Because any fluctuation of the laser intensity is likely to experience some recurrence after one round trip of the radiation in the cavity, the RF spectrum of the output of the FSF laser may show a set of spectral components which are separated in frequency by the free-spectral range  $\Delta_{FSR}$ of the cavity.  Such components separated by $\Delta_{FSR}$ are schematically shown as long thick vertical lines in Fig. \ref{fig10}.

When the distance of the object, and thus $\gamma_c T$, increases the spectral components shown as short vertical dashed lines in Fig. \ref{fig10}  decrease in frequency while the frequency of the other components shown as short vertical solid lines in Fig. \ref{fig10} increase. Each one of these components carries the information about  $\gamma_c T$. The structure shown in Fig. \ref{fig10} is repeated within neighboring frequency intervals of bandwidth  $\Delta_{FSR}$. The following discussion deals only with one of the spectral regions, for instance with the one including the RF components in the range $0 <\Omega/\Delta_{FRS}< 1$. Electronic filters assure that spectral components outside this spectral range are not detected.  In the case shown it happens that  $\Omega^+_{q-1}< \Omega^-_p$. However, depending on the value of   $\gamma_c T$, we could also have  $\Omega^+_{q-1}> \Omega^-_p$. The frequencies of these two components are symmetric about the middle of the given spectral range. For frequency counting to successfully determine $T$ (and thus the distance $  z_{obj}$) only \emph{one} single spectral component is allowed in the spectral range in which the detector plus subsequent electronics has nonzero sensitivity. In order to avoid low frequency noise it is best to use an electronic filter which restricts detection to the higher-frequency part of width  $\Delta_{FSR}/2$, shown in Fig. \ref{fig10} as shaded area.

We will now discuss the straight forward procedure for (i) the determination of the order $p$ or $q$, resolving the ambiguity issue and (ii) how to deal with the situation when $ \gamma T \approx  M  \Delta_{FSR}/2 $ ($M $ is an integer number), i.e. when the spectral components $\Omega^+_{q-1}$ or $\Omega^-_p$    fall near the edge of the sensitive range set by the electronic filter. The latter (problematic) situation is easily overcome by installing a second reference path. The light returned from this additional reference surface is superimposed with a fraction of the light returned from the object on a second detector. The length of the second reference path is adjusted such that the spectral components   are shifted by about an odd integer of $\Delta_{FSR}/2$. With this arrangement, there is always one detector for which the relevant spectral feature is not close the edge of the filtered  bandwidth.

The determination of the order $p$ or $q$ is easily done by repeating a given measurement with a slightly different AOFS frequency $\Delta$. From Eq. (\ref{eq11}) and recalling
that $\gamma_c = \Delta /\tau$
we determine the variation   $\delta\Omega^{\pm}$ of $\Omega^{ \pm}$  when the frequency shift  $\Delta$ is changed by $\delta \Delta$
and arrive for $\Omega^+$ at
\be
      q =  \frac{\Omega^+}{\Delta_{FSR}}\left(1 +\frac{\Delta}{\Omega^{+}}  \frac{\delta\Omega^{+}}{\delta \Delta}\right),
      \label{ambq}
      \ee
      and accordingly for $\Omega^-$ and the order p, starting from Eq. (\ref{eq12}).
Because  $\Delta$ and  $\Delta_{FSR}$ are known, measuring  $\Omega^{\pm}$  and  $\delta\Omega^{\pm}/\delta \Delta$    will lead to the value of the wanted order. Experiments have confirmed that a variation of $\delta\Delta/\Delta\leq  10^{-3} $ suffices to reliably determine the order $p$ or  $q$. The sign of  $\delta\Omega^{\pm}/\delta \Delta$   will   reveal whether we detect a component of type  $\Omega^+$ or  $\Omega^-$.

We conclude this discussing by determining the unambiguity range $z_{ua}$  for our experimental conditions.  As discussed above the determination of   $ z_{obj}$ is unambigious when  $\Omega = \gamma_c T \leq \Delta_{FSR}/2$, what leads with    $T = 2    z_{obj}/ c$ to
\be
 z_{ua}  = \frac{c}{4\Delta}.
\ee
We also note    an alternative approach for resolving the ambiguity issue   used  in \cite{PIQ13}, an approach which is specific to the  FSF-laser system developed for that work. Because that laser radiation involves pulses, the distance is measured in a first step sufficiently accurate by the time-of-flight    method to determine the integer number $p$ or $q$.

\subsection{Experiments}
\label{sec:3c}

\begin{figure}[h]
\centerline{ \resizebox{0.50\textwidth}{!}{%
 \includegraphics[angle=90]{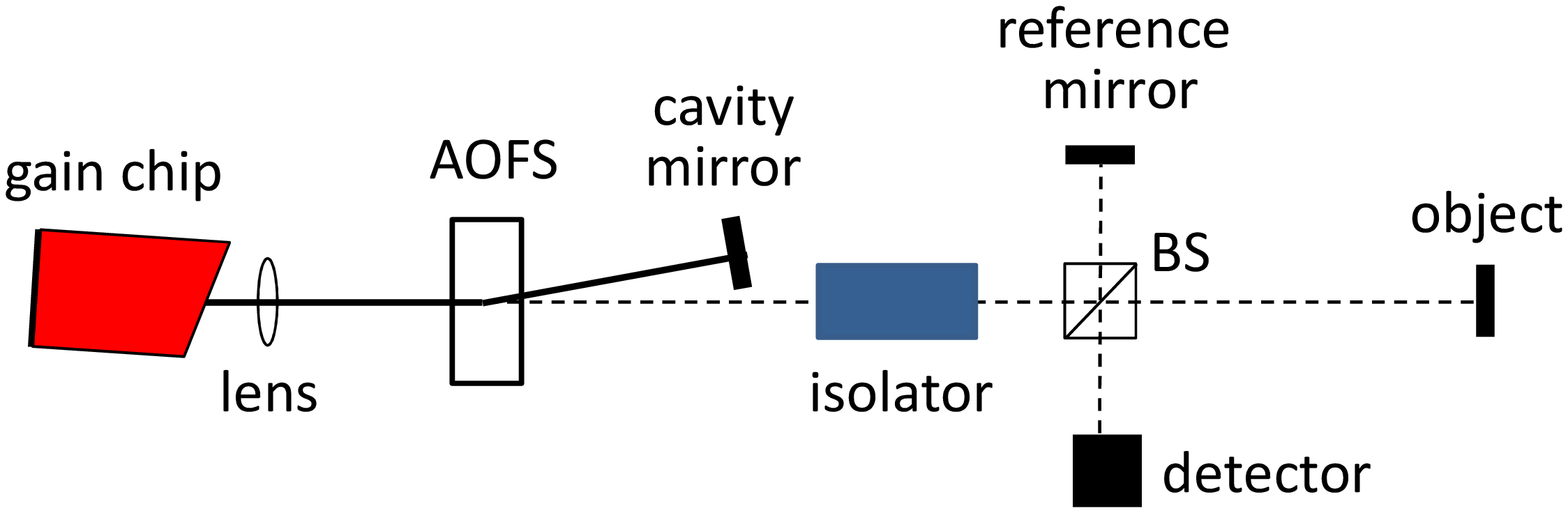}%
 }}
 \caption{ Schematics of the linear-cavity FSF-laser setup, with a gain chip, a collimating lens, an acousto-optic frequency shifter (AOFS) and an interferometer with 50:50 beam splitter (BS), a reference mirror, the object and the detector. An optical isolator prevents feedback into the laser.  For recording the data shown in Fig. \ref{fig12} and \ref{fig13}, the cavity mirror was replaced by a diffraction grating and the zero-order reflection of that grating used as output beam. }
 \label{fig11}
 \end{figure}

The experimental set-up is shown in Fig. \ref{fig11}.   The semiconductor gain chip  (COVEGA TFP780A) working at 780~nm  has one tilted facet with antireflection coating and one normal facet coated for high reflection. This gain medium and the AOFS  with $\Delta_{AOFS}= 80$~MHz frequency shift are placed in a short linear cavity with free-spectral range $\Delta_{FSR}=1.3$~GHz.
\vskip 2 cm

The cavity is closed via the frequency-shifted first diffraction order of the AOFS. Thus, the net frequency-shift  per round trip is $\Delta=2\Delta_{AOFS} = 160$~MHz, leading to a "chirp rate" of  $\gamma_c = 2.1 \cdot 10^{17}$ Hz/s. Typically, the undiffracted beam of the latter with a power of a few mW is used for the measurements.\Black{}     A 50:50 beam-splitter cube (BS) sends this beam to both, the reference surface and the object, some 8~cm away. The light, returning from these two surfaces, is recombined at the same BS and sent to the detector. The frequency of the oscillating component of  the detector output is measured by frequency counting (Tektronix FCA3003).
\begin{figure}[h]
 \centerline{\includegraphics[width=60mm,angle=90]{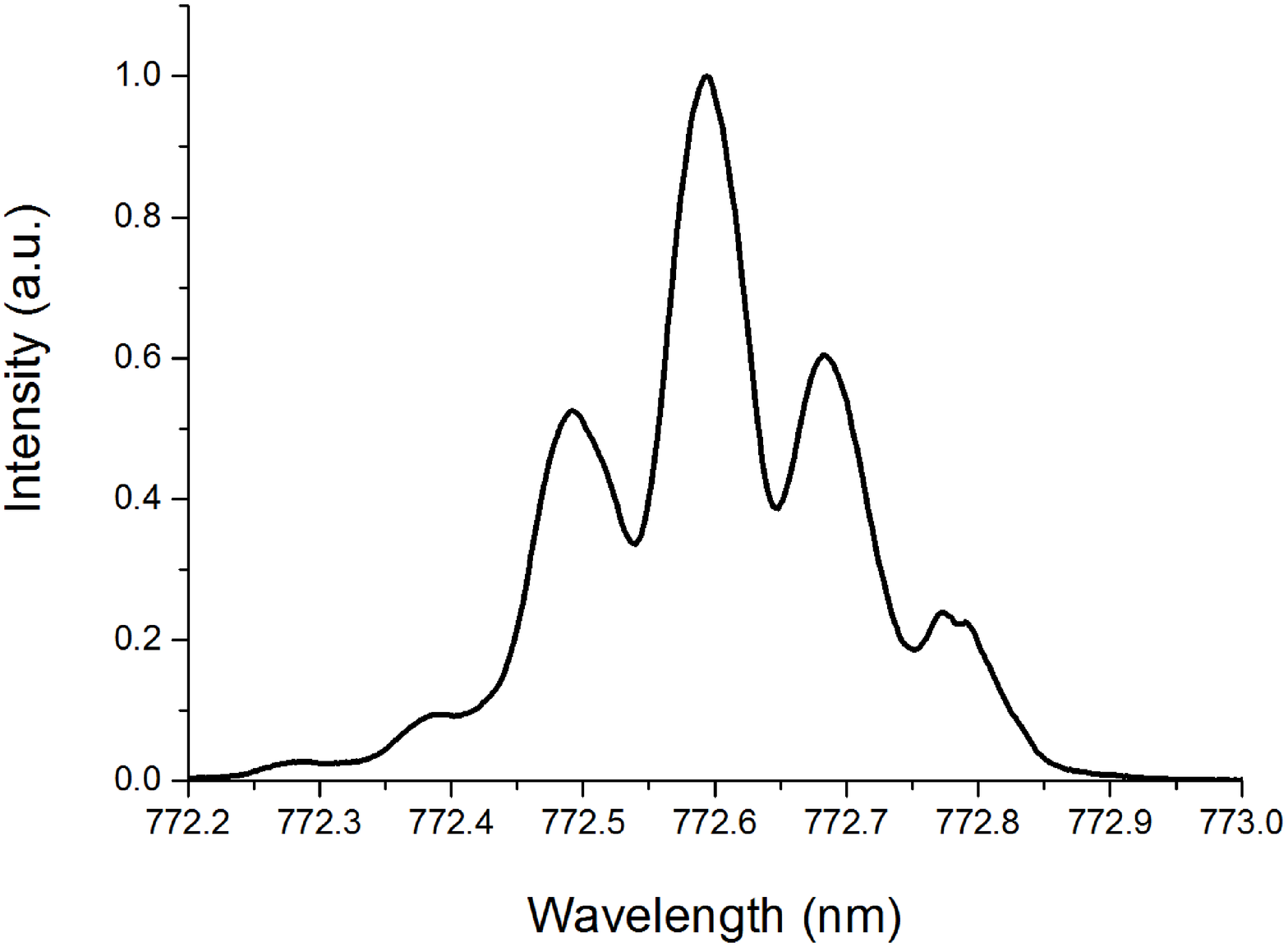}}%
 \caption{Optical spectrum of the   linear-cavity semiconductor FSF-laser  radiation.
 }
 \label{optspectr}
 \end{figure}

Figure \ref{optspectr} shows    a  spectrum of the   linear-cavity semiconductor FSF-laser  radiation with its central wavelength near   $\lambda_c = 772.6$~nm and an overall width   of 0.4~nm.  The residual reflection on the output facet of the gain chip leads to the mode structure  seen in Fig. \ref{optspectr}, with the separation  $\Delta_{chip}$ of the peaks given by the dimension of the gain chip. The individual spectral components of width  $\leq 0.08$~nm  are mutually incoherent. Therefore, the bandwidth which determines the accuracy in a ranging experiment is relatively small.

 Experiments are typically done with an output power of $P = 10$~mW.
Because of the low output power   and small effective bandwidth of the optical spectrum,  the data shown below serve as proof of principle rather than as convincing documentation of the full potential of this approach. For practical ranging application  power amplification is needed, e.g. by a tapered amplifier.  Amplification does not modify the essential spectral properties of the radiation emitted by the FSF laser. Furthermore, suitable modification of the gain chips is likely to lead to a   larger coherence bandwidth and thus to a   higher accuracy of ranging results.
\subsection{Results}
\label{sec:3d}

\begin{figure}[h]
 \centerline{\includegraphics[width=60mm,angle=90]{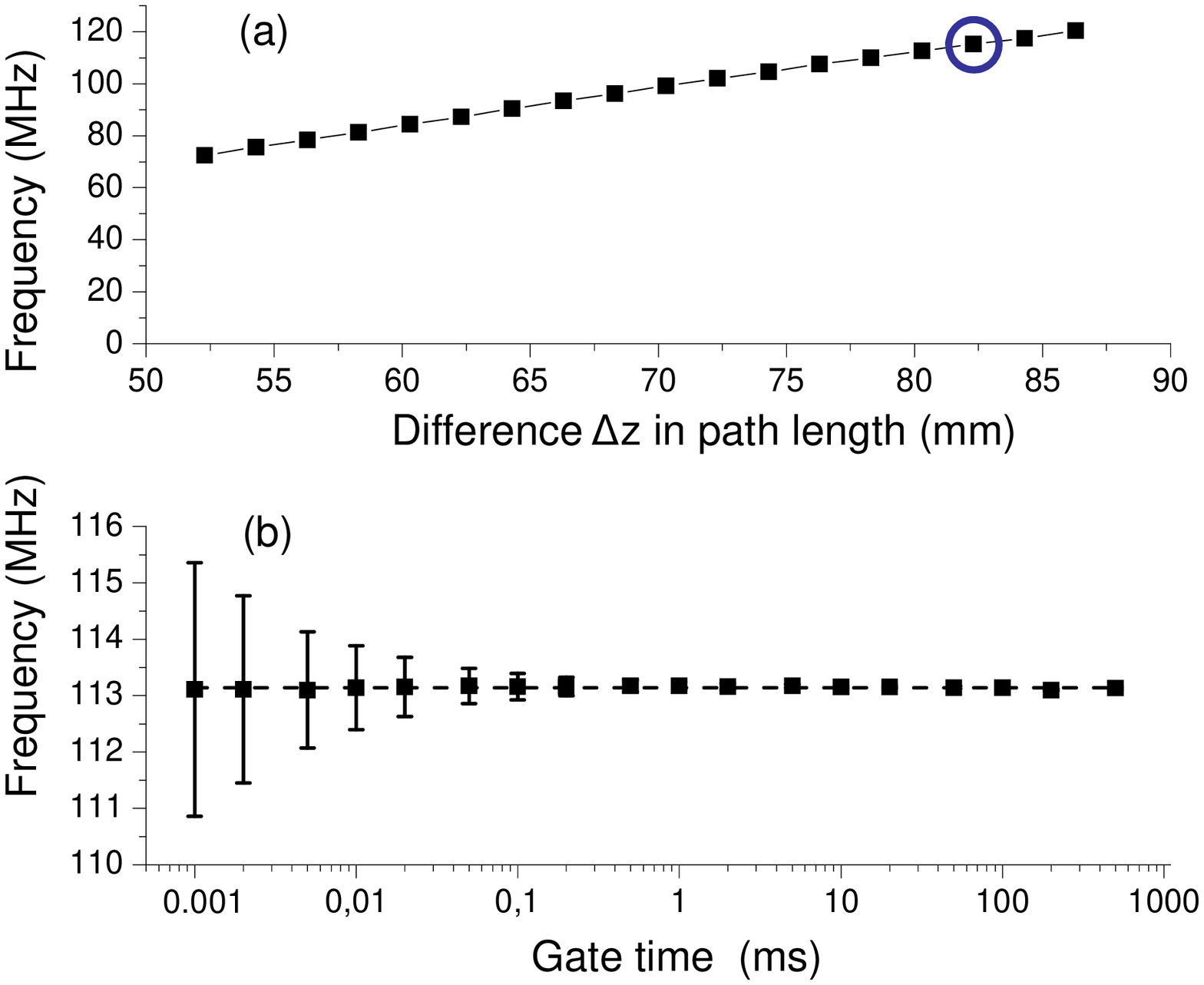}}%
 \caption{  (a) The variation of the center frequency of the beat signals as a function of the distance $\Delta z$. This variation is consistent with the "chirp rate" $\gamma_c=2.1 \cdot 10^{17}$~Hz/s, given by the AOFS frequency and the cavity roundtrip time. (b) The variation of the $1\sigma$-value of the distribution of results from 10$^3$ frequency measurements (taken at the distance marked by the circle in frame (a)) with gate time during which oscillation periods are counted. It is a most important feature that the mean of the measurement does not change when the gate time is varied over nearly six orders of magnitude.
 }
 \vskip 2 cm
 \label{fig12}
 \end{figure}

\begin{figure}[h]
 \resizebox{0.45\textwidth}{!}{%
 \includegraphics[angle=90]{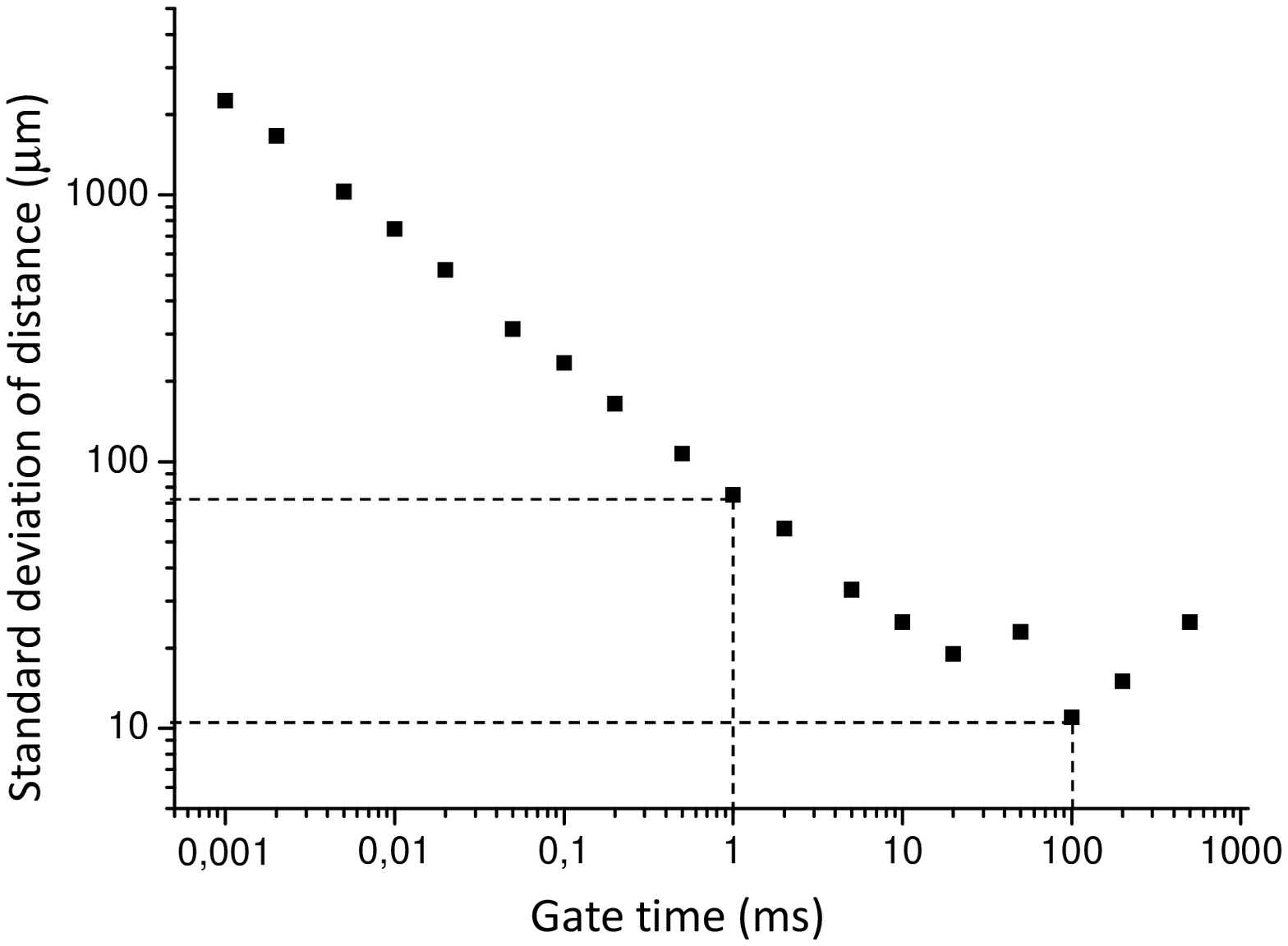}%
 }
 \caption{ Variation of the standard deviation $\sigma$ of distance measurement. We find  $\sigma=2$~mm for a gate time of 1~$\mu$s and $\sigma= 80$~$\mu$m for a gate time of 1~ms. For a gate time $> 10$~ms the limit of $\sigma \geq 10$~$\mu$m is approached.
 }
 \label{fig13}
 \end{figure}

The beat frequency, $\Omega_{beat}$, is measured over a range of about 30~mm as shown in Fig. \ref{fig12}(a).   The variation of the accuracy of the measurement for the fixed distance of $\Delta z = 83$~mm with the gate time during which the frequency is measured by counting oscillation periods   is shown in Fig. \ref{fig12}(b) and  Fig. \ref{fig13}. Each entry is based on 10$^3$ individual runs. Fig. \ref{fig12}(b) shows the mean frequency and $1\sigma$-width of the distribution of recorded frequencies, while Fig. \ref{fig13} shows the data of \ref{fig12}(b) converted to the $1\sigma$-width of the related distribution of readings for the distance.

 An important observation from Fig. \ref{fig12}(b) is the fact that the mean value of the frequency measurement (for fixed distance), does not change when the gate time is varied over nearly six orders of magnitude. The data shown in Fig. \ref{fig13} confirm the typical trade-off between accuracy and data collection time. When the gate time is  1~$\mu$s (allowing a measurement rate of the order of 1~MHz) the uncertainty of a measurement is 2~mm. That uncertainty shrinks to about 75~$\mu$m when the gate time is 1~ms  (allowing a measurement rate of the order of 1~kHz). The accuracy continues to improve with further increase of the gate duration  and approaches the level of  $\sigma = 10$~$\mu$m.  The influence of systematic errors becomes relevant for gate times longer than 100~ms.

\section{SUMMARY AND OUTLOOK}
\label{sec:4}
We have presented results based on powerful modifications of schemes for ranging using FSF-laser radiation, characterized through (i) the implementation of a novel scheme with injection of phase-modulated single-mode radiation into the cavity for accurate (order of  $\mu$m) measurements or (ii) by using frequency counting for fast measurements (order of MHz rate). For (i) we used doped fibers as gain medium exploiting   the large gain bandwidth of the laser transitions leading to the large bandwidth (about 5 nm) of the generated frequency comb  with smooth  envelope.  Such a laser is robust and compact and thus suitable for operation in an  industrial environment.
For (ii) we used a semiconductor gain medium  with the high linear gain and saturation power.  Such a FSF laser  can be realized with a short cavity providing short round-trip time ($\leq 1$~ns) necessary for  high-speed ranging.

Equations (1) and (13) reveal a striking similarity of the FMCW-Ladar technique with the FSF-approach. Although the physics behind the signal generation is distinctly different, it is in both cases a parameter $\gamma_ c$ which maps "distance" onto "frequency". In FMCW-Ladar $\gamma_c$ characterizes the chirp of the radiation with a periodic, saw tooth like, change of the frequency. For precision of the measurements  $\gamma_c$ must be constant, meaning that efforts are needed to assure that the chirp is strictly linear across the entire bandwidth which is covered. In FSF ranging  $\gamma_c$ is typically (much) larger, meaning that a given distance  $z_{obj}$ is mapped onto a (much) larger frequency. Furthermore, because in FSF ranging  $\gamma_c$ is given by well controlled parameters (the frequency shift $\Delta$  induced by the an acousto-optic frequency shifter and the cavity round-trip time  $\tau$) it is straight forward to keep it constant to very high precision. Finally, because the seeding (either by the injected radiation or by spontaneous emission within the gain medium) occurs continuously followed by frequency shift across the gain profile, the radiation is truly continuous.

An obvious next step is to go beyond the current arrangement with circular focus and a single detector and use, for even higher speed of scanning   surfaces of objects, a cylindrical focus and a linear array of detectors or even a planar array of detectors when the entire surface of the object surface is illuminated by the FSF radiation.

\section{ACKNOWLEDGMENT}

We acknowledge support from the German "Bundesministerium f\"{u}r Bildung und Forschung" (BMBF) under the projects numbered 13-N-9345 and 13-N-9346.  K.B. acknowledges additional support from the research center OPTIMAS of the state of Rhineland-Palatinate. We also thank B.W.Shore for carefully reading the manuscript.

\end{document}